\pdfoutput=1
\documentclass[english,letter,oldversion]{aa}
\usepackage[T1]{fontenc}
\usepackage[latin1]{inputenc}
\usepackage{amsmath}
\usepackage{graphicx}
\usepackage{amssymb}

\makeatletter

\providecommand{\tabularnewline}{\\}

\usepackage{txfonts}
\usepackage{natbib}
\bibpunct{(}{)}{;}{a}{}{,}
\def\cite{\citep}

\usepackage{babel}
\makeatother
\begin{document}
\authorrunning{M. Zatloukal et al.}

\titlerunning{Distant galaxy clusters in the COSMOS field found by HIROCS}

\title{Distant galaxy clusters in the COSMOS field found by HIROCS%
\thanks{Based on observations collected at the Centro Astron\'{o}mico Hispano
Alem\'{a}n (CAHA) at Calar Alto, operated jointly by the Max-Planck-Institut
f\"{u}r Astronomie and the Instituto de Astrof\'{i}sica de Andaluc\'{i}a
(CSIC)%
}}

\author{M. Zatloukal\inst{1}\and H.-J. R\"{o}ser\inst{1}\and C. Wolf\inst{2}\and
H. Hippelein \inst{1}\and S. Falter\inst{3}}

\institute{Max-Planck-Institut f\"{u}r Astronomie, K\"{o}nigstuhl 17, D-69117
Heidelberg, Germany\and University of Oxford, Department of physics,
Denys Wilkinson Bldg., Keble Road, Oxford OX1 3RH, United Kingdom\and
I. Physikalisches Institut, Universit\"{a}t zu K\"{o}ln, Z\"{u}lpicherstr.
77, D-50937 K\"{o}ln, Germany}

\offprints{M. Zatloukal, \email{Zatloukal@mpia.de}}

\date{Received 12 June 2007; Accepted 24 August 2007}

\abstract{We present the first high-redshift gal axy cluster candidate sample
from the HIROCS survey found in the COSMOS field. It results from
a combination of public COSMOS with proprietary H-band data on a $0.66\,\square^{\circ}$
part of the COSMOS field and comprises 12 candidates in the redshift
range $1.23\le z\le1.55$. We find an increasing fraction of blue
cluster members with increasing redshift. Many of the blue and even
some of the reddest member galaxies exhibit disturbed morphologies
as well as signs of interaction. \keywords{surveys - galaxies: clusters:
general - galaxies: high-redshift - galaxies: photometry}}

\maketitle

\section{Introduction}

Galaxy clusters provide ideal laboratories to study galaxy evolution.
Furthermore, they can also be used to constrain cosmological parameters
\cite{2005RvMP...77..207V} as well as theoretical models of environmental
effects for which there is growing evidence \cite{2007arXiv0704.2418K}.
Only few galaxy clusters at $z\ge1$ are known today, most of them
stemming from X-ray surveys \cite{2005ApJ...623L..85M,2006ApJ...646L..13S}.
These detect clusters with a hot intra-cluster medium, as will future
surveys based on the Sunyaev-Zel'dovich effect \cite{2001MNRAS.328..783K}.
Optical selection of clusters provides a complementary sample \cite{2004MNRAS.348..551G},
tracing stellar light. At $z\approx1.1$, where the $4000\,\mbox{\AA}$
break gets shifted to the near-infrared, optical surveys need to be
supplemented with at least one near-IR band. With the advent of wide-field
near-IR cameras such as OMEGA2000 on Calar Alto, wide-area imaging
surveys of $z\ge1$ clusters have become feasible \cite{2006ApJ...639..816E,2006astro.ph..4426L}.
A number of optical / near-IR surveys is now beginning to find clusters
in this redshift range \cite{2005ApJ...634L.129S,2006MNRAS.373L..26V}.

In this letter we use proprietary $H$-band data in combination with
COSMOS public data \cite{2006astro.ph.12384S} to find clusters in
the redshift range $1.2\le z\le1.6$ in the COSMOS field. Our detection
method employs photometric redshifts to estimate the distances and
finds clusters as overdensities in 3D galaxy distribution. Hence,
not all our candidates need to be virialized. Some of them may also
be clusters in the state of formation. Throughout this letter we assume
$H_{0}=70\,\mbox{km s}^{-1}\mbox{Mpc}^{-1}$, $\Omega_{m}=0.3$ and
$\Omega_{\Lambda}=0.7$. All magnitudes quoted are in the Vega system.

\section{Data}

HIROCS, the Heidelberg InfraRed / Optical Cluster Survey, aims at
finding galaxy clusters at $0.5\le z\le1.5$ by means of photometric
redshifts. The survey area is $11\,\square\degr$ in four fields,
one of them including the COSMOS field. A detailed description of
the survey will be found in R\"{o}ser \emph{}et al. (in preparation).

In the COSMOS field, HIROCS uses the public data in $u^{*}$, $B_{j}$,
$g+$, $V_{j}$, $r+$, $i+$, $NB816$ and $z+$ \cite{2007arXiv0704.2430C}.
$K_{s}$ is used only as a supplement since it is very shallow compared
to the other bands. We use the spatially psf homogenized matched\_psf
mosaics with PSFs from $0.9\arcsec$ to $1.6\arcsec$ for the individual
filters. The data come calibrated in units of nJy with information
about errors and bad pixels stored in separate RMS images. Morphological
information is provided by HST ACS $F814W$ data \cite{2007astro.ph..3095K}.

As a complement to the optical COSMOS data we use proprietary $H$-band
data taken with OMEGA2000 at Calar Alto Observatory in a joint effort
of the large extragalactic survey projects ALHAMBRA \cite{2005astro.ph..4545M},
MUNICS-deep (Goranova et al., in preparation) and HIROCS. The COSMOS
field will be covered by 25 OMEGA2000 pointings of $15.4\arcmin\times15.4\arcmin$
with a targeted depth of 21 mag ($5\sigma$) and seeing $<1.6\arcsec$.
Currently, the $H$-band data cover $1.3\,\square\degr$ with a non-uniform
depth. For this letter we use the deepest $0.66\,\square\degr$ which
reach 21.4 mag ($3\sigma$) or deeper, see Fig.\,\ref{fig:Overdenstiy}.

\section{Data reduction and analysis}

The OMEGA2000 reduction pipeline \cite{2007.Fassbender} was used
to reduce the $H$-band data applying object masking for sky determination
and image stacking using relative weights and fractional pixel offsets.
Flatfields are derived from the science data. Bad pixel and cosmics
with a difference of $>3\sigma$ to a median image are replaced by
the scaled median value. The images are astrometrically aligned using
point sources from the 2MASS catalogue \cite{2006AJ....131.1163S}.

Our final object catalog contains all objects from all the bands regardless
of their colour with positions as input for the photometry. We use
\textsc{SExtractor} \cite{1996A&AS..117..393B} to extract an object
list from the coadded images for each filter with a minimum S/N threshold
of 5, omitting the shallow $K_{s}$ images. Spurious detections, \emph{e.g.}
in the halos around bright stars, are removed manually. Next, an artificial
image containing all objects from the different \textsc{SExtractor}
tables is created. For each object, a Gaussian is added into this
image with the S/N as amplitude and the shape as measured. The final
catalog is created by running \textsc{SExtractor} on this artificial
image with a special setup for noise-free data. It reaches a mean
positional accuracy of $\sim0.1\arcsec$.

We use \textsc{MPIAPHOT} to derive instrumental magnitudes as weighted
sums over the image area \cite{1991A&A...252..458R}. The width of
the weighting function is adjusted such that the measurements refer
to a common Gaussian {}``aperture'' of $1.8\arcsec$ for all images:
$\sigma_{{\rm common}}^{2}=\sigma_{{\rm weight}}^{2}+\sigma_{{\rm seeing}}^{2}$,
equivalent to the beam in aperture synthesis radio astronomy. Photometry
in the COSMOS bands is done on the coadded frames. The flux errors
are measured using the same parameter files on variance frames created
by squaring the flux of the RMS frames. $H$-band photometry is done
on subsums of 5 one-minute exposures coadded using relative weights
and fractional pixel offsets.

The $H$-band needs a mosaic correction of the instrumental magnitudes
due to varying observing conditions. Each of the 25 pointings is calibrated
using 2MASS stars brighter than $H=14.5$. We get a mean zeropoint
error of $0.008$ with an RMS of $0.07$ measured across the whole
field. Fine tuning of the photometric calibration zeropoint, especially
to relate the optical COSMOS magnitudes to the infrared magnitudes,
is done by applying a stellar main sequence colour shifting technique
similar to the one used by \citet{2001A&A...365..681W}. 

Our object catalog comprises $32\,798$ objects with $H<21.4$ to
be subject to multi-colour classification. Due to the deep COSMOS
optical data, all objects have photometric redshifts assigned and
our catalog is essentially complete down to $H=21.4$.

\begin{figure}
\begin{centering}
\includegraphics[width=1\columnwidth]{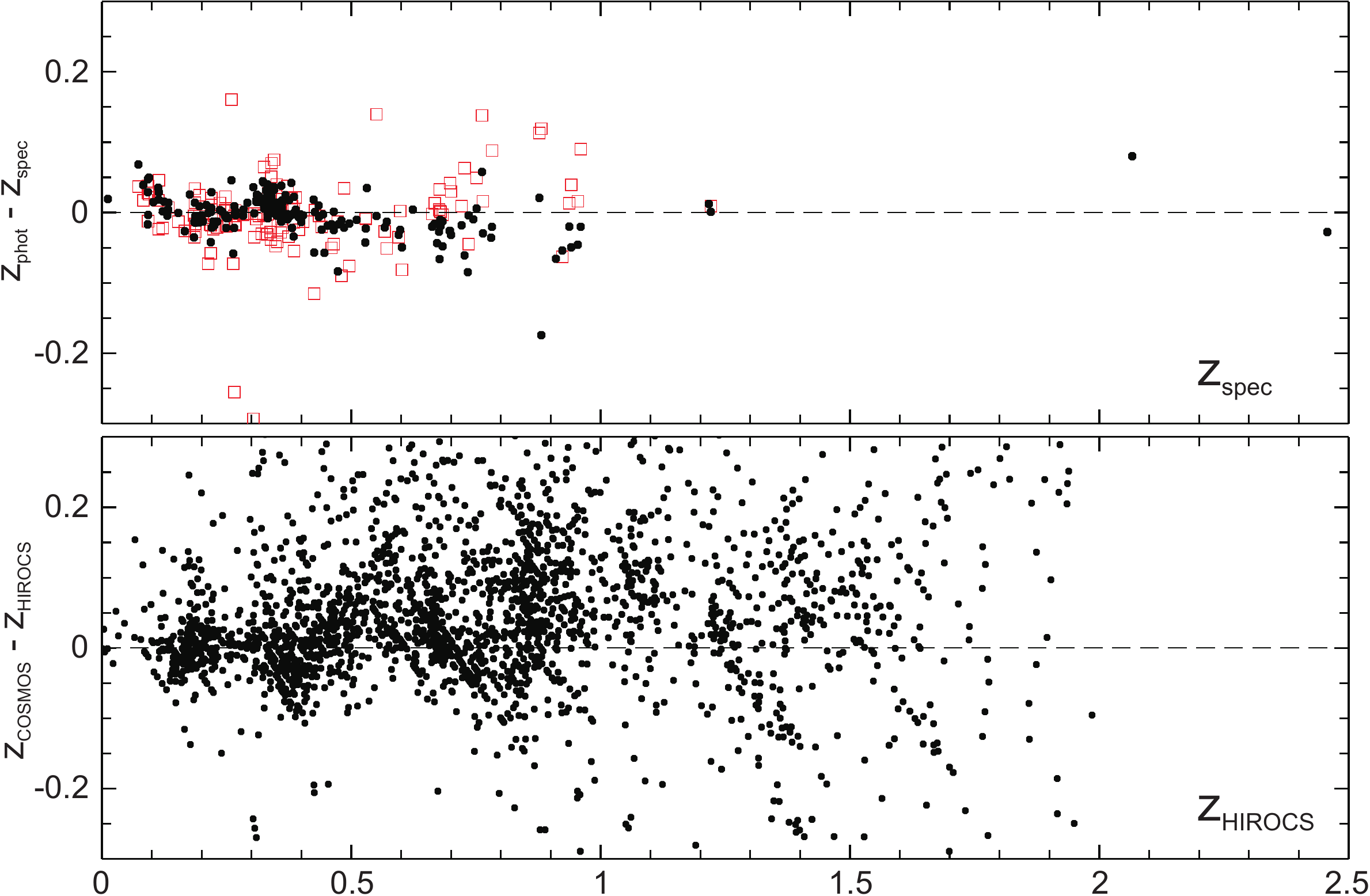}
\par\end{centering}

\caption{\label{fig:PhotzQual}Upper panel: Comparison of spectroscopic and
photometric redshifts for HIROCS (filled cirles) and the COSMOS catalog
(open squares). Lower panel: Comparison of HIROCS and COSMOS photometric
redshifts up to $z=2$.}
\end{figure}

The multi-colour classification and photometric redshift estimation
is done using the template-based code developed for CADIS and COMBO-17
by \citet{2001A&A...365..681W}. The quality of the photometric redshifts
is demonstrated in Fig.\,\ref{fig:PhotzQual}. We use 140 spectroscopic
redshifts from \citet{2006ApJ...644..100P} leaving out the QSOs and
a sample of 48 zCOSMOS \cite{2006astro.ph.12291L} objects from masks
obtained from the ESO archive (Gabasch et al., in preparation) to
compare our photometric redshifts with. Both samples are fitted equally
well by our code. A Gaussian fit to the $\Delta z/(1+z)$ distribution
shows an offset of $0.010$ and scatter of $\sigma=0.028$ for $z<1$.
We get about 2\% catastrophic outliers with $\left|\Delta z/(1+z)\right|>3\sigma=0.084$.
This accuracy is a lower limit at bright magnitudes, while larger
deviations are expected for fainter objects. We use a combination
of the internally expected error and a lower limit that increases
with magnitude and scales with $(1+z)$ as a more conservative error
estimate. A $z=24^{{\rm mag}}$ galaxy at $z=1.4$ thus has a median
error of $\sigma_{z}\sim0.30$. 

A cross-check shows our photo-zs to be in good agreement with the
spectroscopic redshifts and the photo-zs from the COSMOS catalog \cite{2006astro.ph.12344M}
below $z=0.4$. Our code seems to underestimate redshifts at $0.4\le z\le1.0$,
see upper panel of Fig.\,\ref{fig:PhotzQual}. At $z\ge0.75$, the
COSMOS catalog redshifts tend to be a little overestimated compared
to the spectroscopic sample, leading to a systematic difference of
$z\approx0.08\ldots0.1$ with respect to our photo-zs. This trend
seems to continue to higher redshifts (see lower panel of Fig.\,\ref{fig:PhotzQual}).

Rest frame magnitudes are computed by convolving a redshifted filter
curve with the best-fitting template spectral energy distribution
(SED) normalizing with the nearest observed band.

\begin{figure}
\begin{centering}
\includegraphics[width=8cm,keepaspectratio]{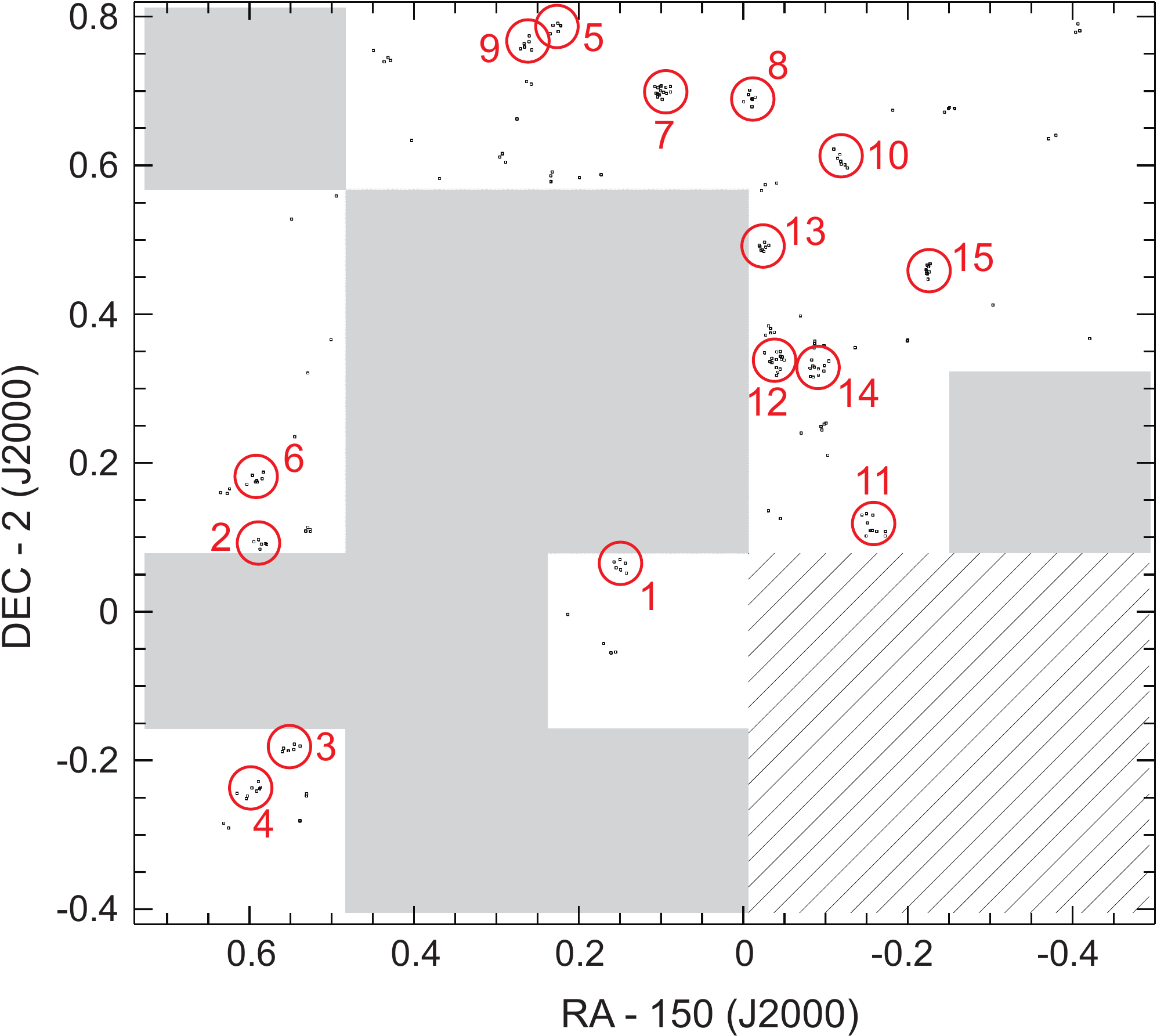}
\par\end{centering}

\caption{\label{fig:Overdenstiy}$H$-band coverage of the COSMOS field and
overdense regions in the redshift range $1.2\le z\le1.6$. In the
dashed region there is no $H$-band data at all, in the light grey
regions it falls short of our target depth of $H=21.4$ ($3\sigma$).
The IDs refer to Table\,\ref{tab:ClusterCatalog}.}
\end{figure}

To find galaxy clusters we use a method based on detecting excess
density in the 3D galaxy distribution (R\"{o}ser et al., in preparation)
by computing the local density for all objects. For each object, the
fractions of the photometric redshift probability distributions $p(z)$
of all its neighbours within $300\,{\rm kpc}$ in projected physical
separation and a $\Delta z$ motivated by the average accuracy of
our photometric redshifts are added. The resulting densities turn
out to be rather insensitive to the exact value of $\Delta z$. Only
objects with $3\sigma$ overdensity compared to the average field
are used to select cluster candidates. All structures with more than
6 overdense members within an aperture of $2\arcmin$ are picked as
cluster candidates from a two-dimensional plot of overdense galaxies
in redshift slices to avoid projection effects. Figure\,\ref{fig:Overdenstiy}
shows such a plot for the redshift range $1.2\le z\le1.6$ with the
cluster candidates highlighted.%
 Monte-Carlo simulations with COMBO-17 data have shown this $3\sigma$
overdensity limit to be rather conservative (Falter et al., in preparation).
Comparison with the density maps of \citet{2006astro.ph.12384S} at
$z<1.1$ shows our overdense objects to trace all density peaks. A
correlation of our HIROCS results with X-ray selected clusters at
$z\ge$0.9 from \citet{2006astro.ph.12360F} in our selection area
shows that four out of six clusters are in common (see Table\,\ref{tab:X-rayClusters}).
The difference in redshift results from the different photo-z codes,
see Fig.\,\ref{fig:PhotzQual}. For X-ray cluster 66 we find no counterpart
in our $H$-selected object list. While this is a low-mass structure,
X-ray cluster 133 is also not detected by our algorithm since it has
no members with a local overdensity $>3\sigma$. Only few galaxies
at its position are visible even if we lower the overdensity cut to
$1\sigma$, despite its mass being above the mass of cluster 102 which
we detect. All these X-ray selected clusters have redshifts $z_{{\rm HIROCS}}<1.2$
and are thus not part of our sample.%
\begin{table}

\caption{\textbf{\label{tab:X-rayClusters}}Correlation of six $z\ge0.9$
X-ray selected clusters from Table\,1 of \citet{2006astro.ph.12360F}.
ID refers to this table, $z_{{\rm XMM}}$ is the redshift given there,
$z_{{\rm HIROCS}}$ is the photometric redshift found by HIROCS. ${\rm M}_{500,{\rm X}}$
is the mass estimate based on X-ray luminosity, $M_{V,{\rm tot}}$the
total rest-frame $V$-band luminosity. $M_{M/L,V}$ is the mass estimate
based on $M_{V,{\rm tot}}$. Masses are given in units of $10^{13}M_{\sun}$.}

\begin{centering}
\begin{tabular}{cccccc}
\hline 
\hline ID&
$z_{{\rm XMM}}$&
$z_{{\rm HIROCS}}$&
${\rm M}_{500,{\rm X}}$&
$M_{V,{\rm tot}}$&
$M_{M/L,V}$ \tabularnewline
\hline 
32&
0.90&
0.79&
$18.64\pm0.52$&
$-25.0$&
$12.0$\tabularnewline
66&
0.95&
-&
$1.81\pm0.45$&
-&
-\tabularnewline
72&
0.90&
0.84&
$5.34\pm0.47$&
$-23.6$&
$3.2$\tabularnewline
102&
1.10&
0.93&
$2.60\pm0.38$&
$-24.5$&
$7.4$\tabularnewline
126&
1.00&
0.86&
$6.87\pm0.69$&
$-24.3$&
$6.1$\tabularnewline
133&
1.15&
-&
$5.68\pm0.64$&
-&
-\tabularnewline
\hline
\end{tabular}
\par\end{centering}
\end{table}

\section{Results and discussion}

\begin{figure}
\begin{centering}
\includegraphics[width=5cm,keepaspectratio]{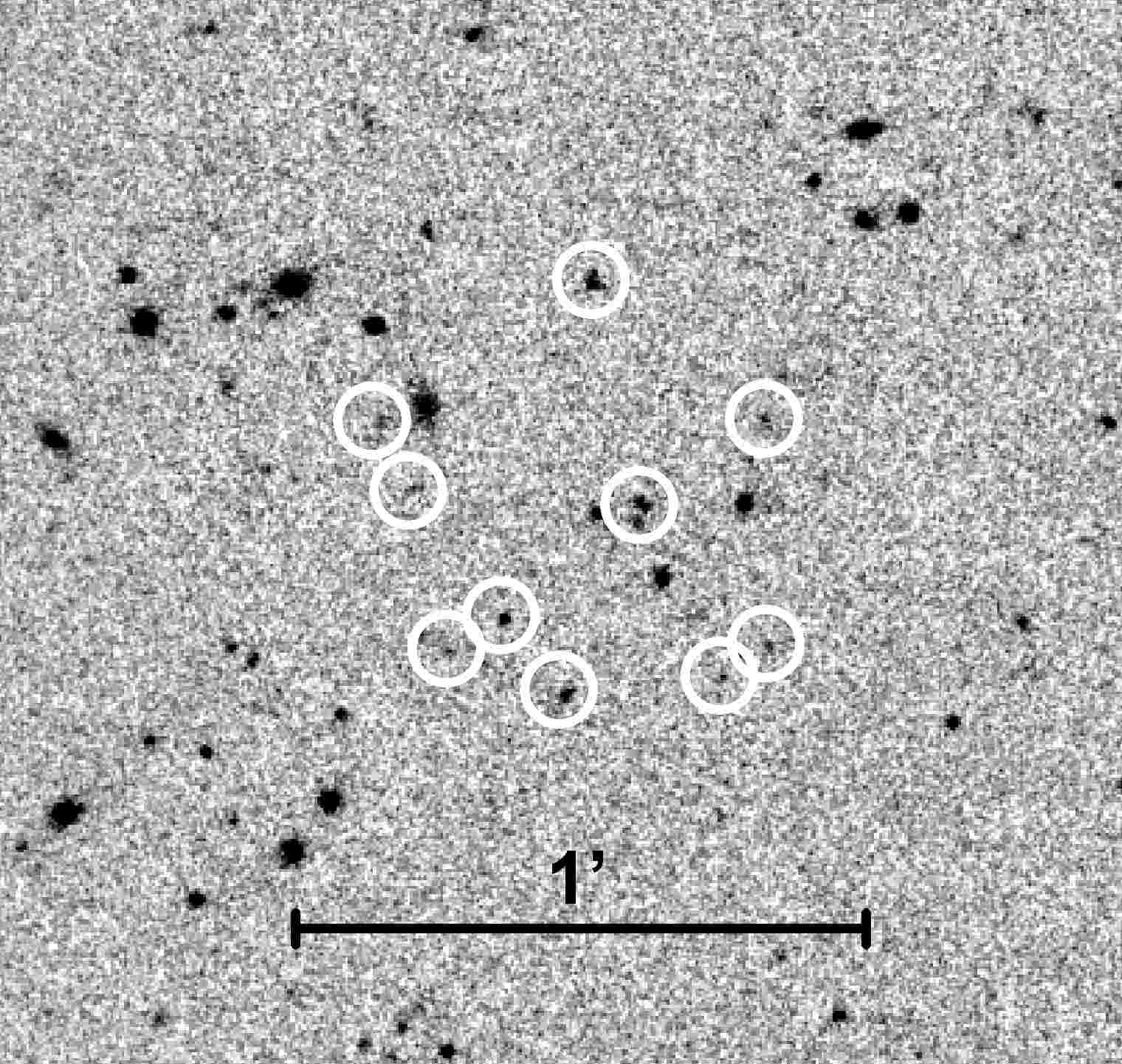}
\par\end{centering}

\caption{\textbf{\label{fig:Cand1H}}$H$-band image of candidate HIROCS$\,$095954.2+022924
(\#13) at $z=1.45$. Member galaxies are marked with white circles. }
\end{figure}

\begin{table*}

\caption{\label{tab:ClusterCatalog}First sample of HIROCS clusters in COSMOS
with $z\ge1.2$. An ID marked with {}``{*}'' means that this candidate
was rejected from the final list, see text for details. The cluster
redshift $z$ is the mean redshift of its members, $\sigma_{z}$ the
error of the mean. $\mbox{N}_{>3\sigma}$ contains the number of members
above the $3\sigma$ overdensity cutoff without correction for field
galaxies, whereas $\mbox{N}_{{\rm tot}}$ gives the total number of
galaxies at the cluster redshift and position regardless of their
local density with the average field density subtracted. They are
selected to be within a circular aperture of $2\times$ the median
distance of the overdense members to the cluster center and to be
in the redshift slice $z_{{\rm cluster}}\pm0.15$. The significance
of $\text{N}_{{\rm tot}}$ with respect to the mean field density
is given in {}``significance''. $M_{V,{\rm BCG}}$ gives the total
restframe $V$-band luminosity of the brightest cluster galaxy, $M_{V,{\rm tot}}$
the field corrected total luminosity in the restframe $V$-band down
to the completeness limit. No correction for completeness is applied
to $M_{V,{\rm tot}}$. $M_{M/L,V}$gives a mass estimate based on
$M_{V,{\rm tot}}$, see text for details. Our 50\% completeness limits
are $M_{V}=-21.4$ at $z=1.3$ and $M_{V}=-21.8$ at $z=1.5$.}

\begin{centering}
\begin{tabular}{cccccccccccc}
\hline 
\hline ID&
Name&
RA {[}$^{\circ}$] &
DEC {[}$^{\circ}$] &
$z$&
$\sigma_{z}$&
$\text{N}_{>3\sigma}$&
$\text{N}_{{\rm tot}}$&
$M_{V,{\rm BCG}}$&
$M_{V,{\rm tot}}$&
significance &
$M_{M/L,V}$\tabularnewline
&
&
(J2000)&
(J2000)&
&
&
&
&
&
&
{[}$\sigma$]&
{[}$10^{13}M_{\sun}$]\tabularnewline
\hline
1{*}&
HIROCS 100035.6+020338&
150.148&
2.061&
1.22&
0.03&
7&
6&
-23.2&
-23.9&
2.4&
2.9\tabularnewline
2&
HIROCS 100220.5+020534&
150.586&
2.093&
1.23&
0.07&
9&
8&
-22.6&
-24.2&
4.5&
4.0\tabularnewline
3&
HIROCS 100212.2+014858&
150.551&
1.816&
1.24&
0.06&
9&
11&
-23.8&
-25.0&
5.3&
7.9\tabularnewline
4&
HIROCS 100223.7+014527&
150.599&
1.758&
1.24&
0.06&
10&
11&
-23.7&
-24.5&
3.5&
5.0\tabularnewline
5&
HIROCS 100054.5+024703&
150.227&
2.784&
1.26&
0.06&
7&
12&
-24.0&
-24.8&
9.0&
6.5\tabularnewline
6&
HIROCS 100221.8+021045&
150.591&
2.179&
1.31&
0.05&
8&
10&
-23.3&
-24.8&
6.4&
6.1\tabularnewline
7{*}&
HIROCS 100023.9+024158&
150.100&
2.699&
1.32&
0.09&
17&
16&
-23.1&
-25.2&
9.1&
8.8\tabularnewline
8&
HIROCS 095957.9+024125&
149.991&
2.690&
1.33&
0.06&
8&
10&
-23.4&
-24.9&
7.3&
6.2\tabularnewline
9&
HIROCS 100103.2+024546&
150.263&
2.763&
1.40&
0.12&
6&
8&
-23.5&
-24.7&
6.6&
4.7\tabularnewline
10&
HIROCS 095931.8+023630&
149.882&
2.608&
1.41&
0.04&
6&
8&
-23.1&
-24.7&
4.4&
4.7\tabularnewline
11{*}&
HIROCS 095922.3+020654&
149.843&
2.115&
1.43&
0.04&
10&
19&
-23.4&
-25.8&
2.5&
13.1\tabularnewline
12&
HIROCS 095950.3+022014&
149.960&
2.337&
1.44&
0.07&
15&
19&
-24.3&
-25.9&
8.4&
14.3\tabularnewline
13&
HIROCS 095954.2+022924&
149.976&
2.490&
1.45&
0.05&
8&
10&
-23.6&
-24.9&
8.0&
5.2\tabularnewline
14&
HIROCS 095938.4+021939&
149.910&
2.328&
1.52&
0.06&
14&
18&
-23.0&
-25.3&
7.7&
6.8\tabularnewline
15&
HIROCS 095906.0+022726&
149.775&
2.457&
1.55&
0.05&
8&
8&
-23.2&
-23.9&
11.4&
1.7\tabularnewline
\hline
\end{tabular}
\par\end{centering}
\end{table*}
The first HIROCS cluster candidate sample for the COSMOS field comprises
15 galaxy cluster candidates with redshifts $1.22\leq z\leq1.55$.
They are listed in Table\,\ref{tab:ClusterCatalog} with their positions
and properties. Their significance with respect to the mean number
density of galaxies in this redshift slice is between $2.4\sigma$
and $11.4\sigma$. Candidates \#1 and \#11 are removed from the final
list due to their low significance, candidate \#7 because it does
not show a clear redshift peak and colour-magnitude plots show it
to be most likely a chance projection of two structures. This leaves
12 candidates in the final list. The number of cluster members is
limited by the depth of the $H$-band data. Figure\,\ref{fig:Cand1H}
shows an $H$-band image of HIROCS$\,$095954.2+022924 (\#13) at $z=1.45$
as an example. Most member galaxies are clustered in a region of about
$1\,\square\arcmin$ on the sky.%

\begin{figure*}
\begin{centering}
\includegraphics[width=18cm,keepaspectratio]{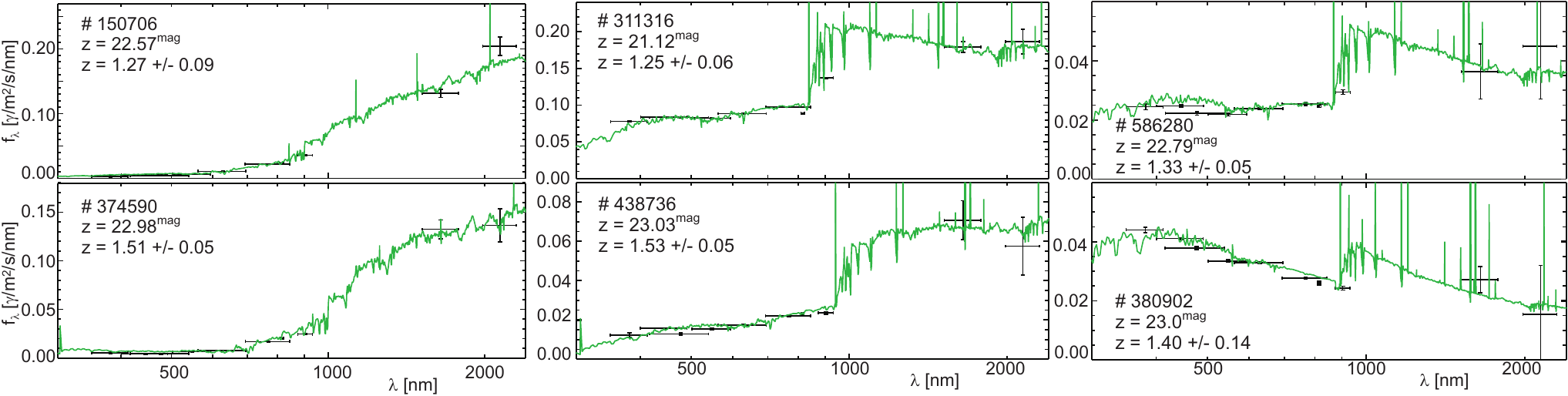}
\par\end{centering}

\caption{\label{fig:SEDs}Representative SEDs of cluster candidate members
in two redshift slices. $z\approx1.25$ in the upper and $z\approx1.5$
in the lower row. The left column shows very red SEDs whereas the
middle and right columns show blue SEDs, indicative of recent or ongoing
star formation. The redshift errors quoted are the results of the
template fitting algorithm.}
\end{figure*}

\begin{figure}
\begin{centering}
\includegraphics[width=0.8\columnwidth]{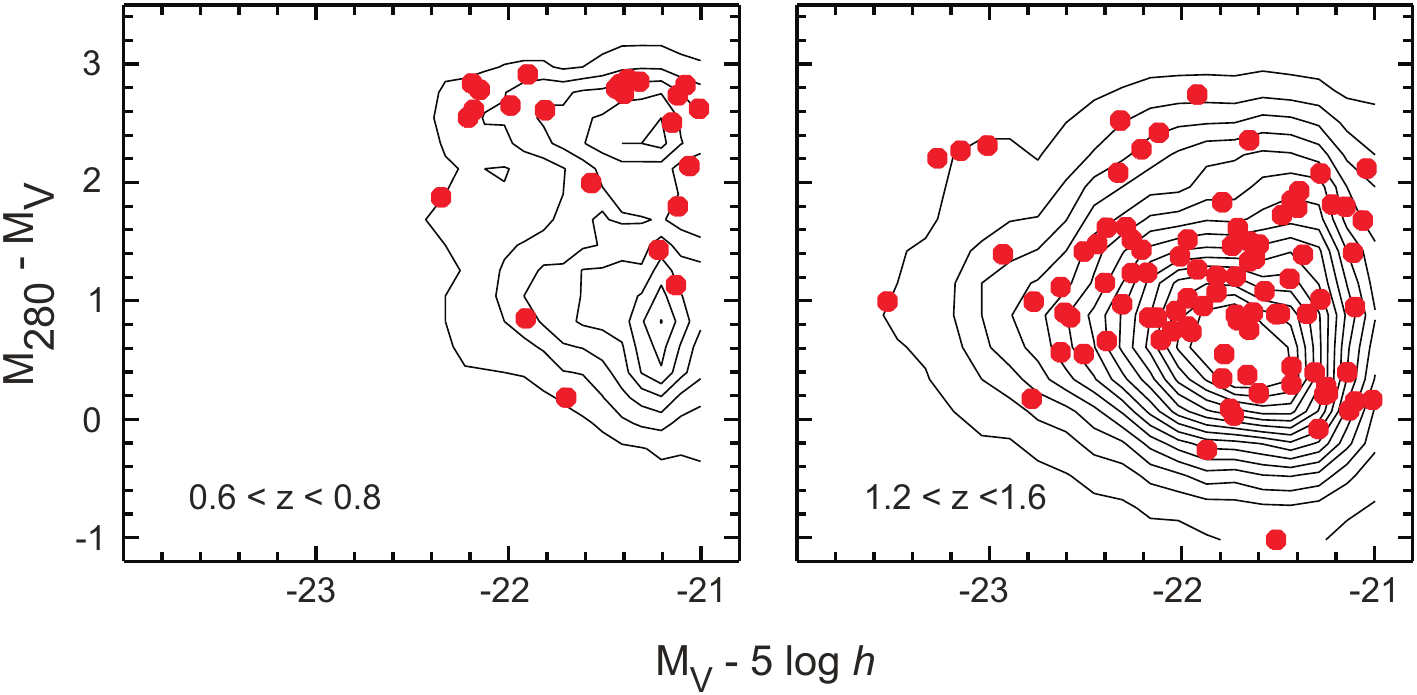}
\par\end{centering}

\caption{\label{fig:RedSequence}The restframe $M_{280}-M_{V}$ colour-magnitude
diagram in the redshift slices $[0.6,\,0.8]$ (left) and $[1.2,\,1.6]$
(right). Field galaxies are indicated as contours. Overplotted in
red are the members of a known large-scale structure at $z\approx0.7$
(left) and the HIROCS cluster candidates from Table\,\ref{tab:ClusterCatalog}
(right).}
\end{figure}

Some representative cluster members' SEDs from the template fitting
process are shown in Fig.\,\ref{fig:SEDs}. Note how the $z$-band
flux drops when moving up in redshift from $z=1.2$ to $z=1.4$ (center
column), placing a secure lower limit on the photometric redshift.
While we observe the very red, passively evolving galaxies expected
to be present in galaxy clusters, there is a remarkably high fraction
of galaxies with blue SEDs. This is also evident in colour-magnitude
diagrams of the cluster candidates. %
 Figure\,\ref{fig:RedSequence} shows restframe $M_{280}-M_{V}$
colour-magnitude diagrams of all galaxies with $H\le21.4$ in the
redshift slices {[}$0.6,\,0.8$] (left) and {[}$1.2,\,1.6$] (right).
Overplotted in red are the members of a rich large-scale structure
at a redshift of $z\approx0.7$ \cite{2007astro.ph..1482G} (left
panel) and members of the cluster candidates given in Table\,\ref{tab:ClusterCatalog}
(right panel). The strength of the colour bimodality decreases with
redshift, as observed for example by \citet{2006astro.ph..9287C}.
The fraction of red galaxies is considerably lower in our high-redshift
candidates than in the $z=0.7$ structure. This could be due to a
Butcher-Oemler type evolution, a colour-density effect or a combination
of both. %

On the HST ACS images many members exhibit disturbed morphologies
and signs of interactions. This is seen for a lot of the blue and
even for some of the reddest galaxies. Note that the term {}``red''
only refers to colour and includes both passively evolving old stellar
populations as well as dust-reddened objects.

The total rest-frame $V$-band luminosity $M_{V,{\rm tot}}$ of the
candidates is in the range $-25.8\le M_{V,{\rm tot}}\le-24.2$ with
the brightest cluster galaxies having rest frame luminosities in the
range $-24.3\le M_{V,{\rm BCG}}\le-22.6$. The brightest of these
are consistent with present-day BCGs \cite{AQ4} when applying a passive
evolution correction factor based on the evolution of $M_{B}^{*}$
from \citet{2004ApJ...608..752B} with $z_{{\rm form}}=3$. Our mass
estimates $M_{M/L,V}$ based on the total rest-frame luminosity are
in good agreement with the X-ray based $M_{500}$ masses, see Table\,\ref{tab:X-rayClusters}.
We assume mass-to-light ratios of 140 ($z=0.9$) and 75 ($z=1.4$)
based on the present-day mass-to-light ratio of 300 \cite{AQ4} and
the correction for evolution described above. Comparison with the
mass estimates of the X-ray selected catalog shows that our algorithm
detects structures down to at least $M_{500,X}=2.6\cdot10^{13}M_{\sun}$,
the mass of X-ray selected cluster 102, at $z=0.95$. Our lowest-mass
$z\ge1.2$ candidate HIROCS 095906.0+022726 (\#15) has a total mass
of $M_{M/L,V}\approx1.7\cdot10^{13}M_{\sun}$. At present we do not
have a completeness limit estimate for our algorithm. Theoretical
models ($\Lambda{\rm CDM},\,\sigma_{8}=0.8$) predict about 12 clusters
per $0.66\,\square\degr$ at $z>1.1$ with $M\ge5\cdot10^{13}M_{\sun}$
\cite[M. Bartelmann priv. comm.]{2002A&A...388..732B}. If we take
$M_{200}=1.7\cdot M_{500}$ \cite{2006astro.ph.12360F} and assume
$M_{500}=M_{M/L,V}$ 11 of our 12 candidates fall into this mass range,
matching the predictions. Note that no contamination or completeness
corrections have been applied. The cluster candidates we find are
comparable in redshift to the most distant spectroscopically confirmed
\cite{2007astro.ph..1787M} and photometrically selected \cite{2007astro.ph..0707.1783}
clusters known today. These results indicate that with the full HIROCS
galaxy cluster sample implications on cosmological parameters and
galaxy evolution will become assessible.

\begin{acknowledgements}
The authors would like to thank the COSMOS team for making their data
publicly accessible, R. Fa{\ss}bender for providing the IR reduction
pipeline and the staff at Calar Alto Observatory for their support.
The data were taken in a joint effort with the MUNICS-deep team (U.
Hopp, MPE Garching) and the ALHAMBRA team (M. Moles, IAA Granada).
It is a pleasure to thank both teams for the kind collaboration. The
authors would also like to thank S. Khochfar for valuable comments
on the draft version of this letter, I. Sakelliou for helpful discussions
on the X-ray data and S. Noll (MPE Garching) for sharing his spectroscopic
comparison sample. This publication makes use of the NASA ADS and
data products from the Two Micron All Sky Survey, as well as of the
NASA/ IPAC Infrared Science Archive. We thank an anonymous referee
for a thorough and very helpful report.

\bibliographystyle{aa}
\bibliography{8063}

\end{acknowledgements}

\end{document}